## Main Manuscript for

# Colimitation of light and nitrogen on algal growth revealed by an array microhabitat platform

Fangchen Liu[1], Larissa Gaul[1], Andrea Giometto[2]* and Mingming Wu[1]*

[1]Department of Biological and Environmental Engineering, Cornell University, Ithaca, NY, USA.

[2]School of Civil and Environmental Engineering, Cornell University, Ithaca, NY, USA

*Mingming Wu

***Email:** mw272@cornell.edu

*Andrea Giometto

***Email:** giometto@cornell.edu

**Author Contributions:** FL and MW created and designed the research project. FL carried out all the experiments and data analysis. LG contributed to the original development of the light gradient platform. AG contributed to mathematical modeling. All authors participated in the writing and discussion of the manuscript.

**Competing Interest Statement:** The authors declare no conflict of interest.

**This PDF file includes:**

Main Text
Figures 1 to 5

## Abstract

Microalgae are key players in the global carbon cycle and emerging producers of biofuels. Algal growth is critically regulated by its complex microenvironment, including nitrogen and phosphorous levels, light intensity, and temperature. Mechanistic understanding of algal growth is important for maintaining a balanced ecosystem at a time of climate change and population expansion, as well as providing essential formulations for optimizing biofuel production. Current mathematical models for algal growth in complex environmental conditions are still in their infancy, due in part to the lack of experimental tools necessary to generate data amenable to theoretical modeling. Here, we present a high throughput microfluidic platform that allows for algal growth with precise control over light intensity and nutrient gradients, while also performing real-time microscopic imaging. We propose a general mathematical model that describes algal growth under multiple physical and chemical environments, which we have validated experimentally. We showed that light and nitrogen colimited the growth of the model alga *Chlamydomonas reinhardtii* following a multiplicative Monod kinetic model. The microfluidic platform presented here can be easily adapted to studies of other photosynthetic micro-organisms, and the algal growth model will be essential for future bioreactor designs and ecological predictions.

## Significance Statement

Mechanistic understanding of the impact of complex environmental parameters on the growth of photosynthetic micro-organisms is critical for maintaining the balance of ecosystems, as well as optimizing the production of clean biofuels. Here, we present the first microfluidic platform for providing well defined physical and chemical gradients to algal cells. Our measurements revealed that light and nitrogen colimit the growth of microalgae, and their effects are well described by a multiplicative Monod kinetic model. This work provides a microfluidic platform for quantitative studies of the growth of photosynthetic micro-organisms, and a general mathematical model for algal cell growth.

## Main Text

## Introduction

Phytoplankton, including microalgae and cyanobacteria, are essential in maintaining the balance of ecosystems contributing to primary production and the carbon cycle. By producing and releasing oxygen as a byproduct of photosynthesis, cyanobacteria are known to be the first oxygen producers on Earth, dating back billions of years ago, and are essential for life on Earth. Disruption of phytoplankton growth can lead to environmental problems such as Harmful Algal Blooms (HABs) (1-3), which deplete threatened water resources. The problem of HABs has been recently exacerbated by climate change, including the effect of warming temperatures and increasing frequency of storms and flooding events (4-6). In contrary to HABs, lipid production from controlled growth of microalgae is a promising avenue for clean alternative bioenergy (7-9). A number of microalgae species have been found to be excellent nutritional food sources for animals as well as humans (10-12). As such, a mechanistic understanding of and the ability to control algal growth is an essential step towards finding solutions for sustainable living.

The biophysical (e.g., light and temperature) and biochemical (e.g., nitrogen and phosphorous) environment critically impacts the growth of microalgae. Traditionally, the environment has been considered as the major selection force acting on genes. Recent studies have revealed important roles played by the environment in ecology, evolution, and development of biological organisms, within the so-called “eco-evo-devo” field (13-17). Phenotypic plasticity of cells can emerge under time-varying environments, which help cells respond quickly to environmental fluctuations (18-20). Also, the passing down of extra-cellular environments could serve as another mechanism for non-genetic inheritance, which could not only affect single cells, but also a community of cells and their interactions (21-25). For example, microorganisms and the interactions among them can shape their environment, a phenomenon known as niche construction, leading to the co-evolution of organisms and their environment (26-29). Despite the importance of cell-environment interactions in algal growth, mechanistic understanding of how complex environments regulate algal growth is limited. This is due to the lack of tools where environmental conditions can be directly controlled while cell growth is simultaneously monitored in real time at a quantitative level.

Microfluidics has emerged to provide well controlled environments for algal cell growth. Algal growth has been studied extensively under single environmental gradients, including nutrients (30-33) and light (34, 35). Due to the complexity of the environment in nature, recent work has started to explore roles of multiple environmental parameters in cell growth. Dual microfluidic chemical gradients have been reported to rapidly screen cell response to multiple stressors or characterize cell responses to stressors under complex environments (36-39). Specific to algal cells, a microfluidic dual chemical gradient generator device revealed the synergistic effect of nutrients (nitrogen and phosphorous) on cell growth (40). Although photosynthesis is an important part of algal cell growth, the quantitative understanding of the effect of light exposure on algal growth is limited, especially under a controlled nutrient environment (41, 42). Nguyen et al. recently presented an important millimeter-scale platform that examined the optimal light intensity and nutrient condition for lipid production of algal cells under controlled light and nutrient condition for cells immobilized in hydrogels (42).

Mathematical modeling of algal growth kinetics, especially in complex environmental conditions, has lacked precise experimental validation. Current algal growth models have been derived from first principles such as mass action kinetics and growth optimality. These models were used to describe observations of natural waters and experiments with large-scale static and continuous cultures for the dependence of algal growth on multiple resources (43-47). However, limitations in the range of substrate concentrations and temporal resolution prevented the validation of these models, limiting our ability to identify the fundamental principles controlling the dependence of algal growth parameters on the concentration of multiple resources. In this spirit, microfluidics could serve as a useful tool to provide well-defined microenvironments and quantitative measurements for the development of accurate growth models for microalgae.

Here, we propose a microfluidic platform that allows us to establish gradients of both light intensity and nutrient concentration directly on the array microhabitats placed on a microscope stage. Key innovations of our platform include that growing *C. reinhardtii* cells can swim freely within microhabitats, closely mimicking its natural habitat, and that the growth dynamics can be continuously monitored in real time instead of end point measurements. We propose a general mathematical model developed alongside experimental data demonstrating that light intensity and nutrient concentration colimit the growth of algal cells following a general multiplicative Monod kinetic model.

## Results and Discussion

### A microfluidic platform for growing photosynthetic microbes under a dual light and nitrogen gradient

A hydrogel-based array microhabitat device in conjunction with a light gradient was used to study algal growth under well-defined physical and chemical conditions (See Fig. 1A). To create a nitrogen gradient, we used a previously developed hydrogel-based chemical gradient generator (40). The device consists of an 8 x 8 array of microhabitats (each with size of 100µm x 100µm x 100µm) flanked by two side channels (400µm W x 200µm H), patterned on a 1 inch x 3 inch size and 1 mm thick agarose gel membrane (Fig. 1A and Fig. S1). Nitrogen-starved cells were seeded in the microhabitats. The nitrogen gradient was established by flowing medium with known nitrogen concentrations and blank buffer through the top and bottom side channels respectively, creating a nutrient gradient at the location of the array microhabitats via molecular diffusion. The time to reach a steady state nutrient gradient was measured to be about 90min (40). The light intensity gradient was generated by modifying the bright field illumination light path of a commercial microscope (Fig. 1A and ref. (48)). Briefly, the light intensity gradient at the location of the array microhabitat was created by placing a half-moon shaped mask directly below the field iris of an Olympus IX81 microscope.

Fig. 1B shows an image of cells growing in microhabitats under light and nitrogen gradients, which were characterized in the following ways. The nitrogen concentration gradient in the dual gradient experiment is shown in Fig. 1C. To characterize the chemical gradient in the array microhabitat, we flowed a solution of a known concentration of a fluorescent dye in the source channel and blank buffer in the sink channel as described previously (40). The steady linear gradient computed from the fluorescence image is shown in Fig. 1C. Here, we assumed the nitrogen concentration in the sink and source channels was 5.3 and 35.3µM respectively, and the concentration of the fluorescence solution was linearly related to the nitrogen concentration. Nitrogen concentrations at the middle line of each row of microhabitats were taken as the concentration of that row. The light intensity was characterized using a light meter and readouts from a CCD camera. Fig. 1D shows an almost linear light intensity profile across the array microhabitat. The maximum PAR value was approximately 45 $\mu mol \cdot m^{-2} \cdot s^{-1}$ and the minimum was 0.1 $\mu mol \cdot m^{-2} \cdot s^{-1}$ (Fig. 1D). PAR values were converted from the grayscale values measured from bright field images of the light intensity gradient captured with a CCD camera as described previously (48). The light intensity of each column was calculated as the average across the width of the microhabitats in that column, as cells were observed to swim freely within single habitats.

To monitor cell growth dynamics, time lapse fluorescence images were taken every four hours during the 7-day experimental period (Fig. 1E). Cell numbers were measured using the fluorescence intensities in each microhabitat, and were used to calculate the growth rates.

Our platform advances the current technology to study photosynthetic microorganisms under well-controlled nutrient and light gradients. The unique feature is the integration of a microfluidic chemical gradient generator together with a microscope-based light gradient generator. To our knowledge, this is the first time that cell growth can be monitored in real time while the cells are subjected to well defined nutrient and light gradient conditions. This platform is particularly suitable for the creation of data driven mathematical growth models for photosynthetic cells that are sensitive to complex microenvironments.

**Algal growth sensitivity to light was enhanced at high nitrogen concentration**

*C. reinhardtii* cells grew very differently under the same light intensity gradient with low versus high nitrogen concentrations (Fig. 2A-B). Cells were starved in medium with 5.3µM nitrogen before loading them into the microhabitats (Materials and Methods). When the cells were provided with a low nitrogen concentration of 5.3µM in the device, no clear response to light was observed (Fig. 2A). In contrast, when 30.8µM of nitrogen was provided uniformly across all the habitats, cells showed a clear increase in growth rate as light intensities increased (Fig. 2B). Fig. 2C-D shows the corresponding growth curves, normalized by the initial cell number that was usually 1-6 cells per habitat due to the random seeding process. Microhabitats without cells were excluded from the analysis. The normalization with respect to the initial cell number revealed a slight growth response to light intensity at 5.3µM nitrogen (Fig. 2C), which was not apparent from the fluorescence images (Fig. 2A). In the presence of 30.8µM nitrogen, the growth curves at different light intensities were clearly spread out (Fig. 2D), indicating that cell growth had higher sensitivity to light at the higher nitrogen concentration.

Light is an important energy source for photosynthetic microbes like microalgae. Energy harnessed from light can be used for the biosynthesis of structural and functional components of the cell, contributing to cell proliferation. Upon the removal of nitrogen, proteins and pigments related to photosynthesis decrease in abundance, including RuBisCO, a key enzyme in the Calvin cycle, and chlorophyll, the pigment for capturing photons, which leads to reduced efficiency of light utilization (49-52). Our results in the microhabitats indeed showed that cell growth was less sensitive to light at low nitrogen concentration, indicating a suppressed photosynthetic capacity. Increased sensitivity to light in the presence of nitrogen as compared to no nitrogen has been shown in previous studies either by -omics response or by photosynthetic functional measurements (49, 50). Here, we showed that growth rate, which results from all cellular processes combined, was more sensitive to light under the high nitrogen concentration.

**Algal growth sensitivity to nitrogen was enhanced under high light intensity**

Higher light intensity, in turn, increased the growth sensitivity of algal cells to nitrogen (Fig. 3). Under lower light intensity (0.1PAR), the cell growth did not vary distinctly across different nitrogen concentrations (Fig. 3A,C). In contrast, at higher light intensity (41.1PAR), cells at lower nitrogen concentrations grew significantly slower than those at higher nitrogen concentrations (Fig. 3B,D). This phenomenon was clearly visible in the fluorescence images (Fig. 3A-B) and was reflected in the growth curves (Fig. 3C-D).

Nitrogen is one of the most important nutrients for algae, contributing to the synthesis of amino acids and proteins. Various forms of nitrogen can be utilized by algal cells including nitrate, nitrite, ammonium, and some organic forms, among which ammonium is the preferred form. Ammonium is transported into cells via the ammonium transporter proteins (AMT family) and is assimilated through the glutamine synthetase-glutamate synthase pathway. Synthesized glutamine and glutamate can be used for the biosynthesis of macromolecules for cell function and proliferation. Increase in light intensity was found to induce responses in various metabolic processes including photosynthesis, as well as amino acid, fatty acid, and nucleotide biosynthesis. Nitrogen metabolism is also known to be affected by light (53-55). Specifically in the case of ammonium utilization, the presence of light contributes to higher activity of glutamate synthases (56). The experimental observation presented here begins to reveal the interactions between nitrogen metabolism and photosynthesis machineries.

**Light intensity and nitrogen concentration synergistically influence the growth rate of algal cells**

Light and nitrogen were found to synergistically promote algal growth when the algal cells were grown in the array microhabitats in the presence of light and nitrogen gradients. Here, the nitrogen gradient was generated by flowing medium with 35.3µM nitrogen in the source channel and medium with 5.3µM nitrogen in the sink channel. The light intensity was provided by the bright field light source of the microscope, ranging from 0 to 45 $\mu mol \cdot m^{-2} \cdot s^{-1}$. The synergistic effect can be seen clearly in Fig. 4A, which shows fluorescence images of cells in the array microhabitats. Therein, the upper right corner represents the habitat with the highest light intensity and highest nitrogen concentration. To understand quantitatively how cell growth depends on light intensity and nitrogen concentration gradients, we calculated growth rates of each microhabitat and displayed them in Fig. 4B. Results from three replicated experiments were shown in Fig. 4B. In all three replicates, the data was consistent, and shows that growth rate was highest at high nitrogen concentration and light intensity.

We noticed an interesting parallel between the cell growth response to dual light and nitrogen gradients (shown in Fig. 4) and our previous work where cell growth response was studied under dual nitrogen and phosphorous gradients (40). Under a single nutrient or light gradient, we used a microfluidic platform to reveal that algal cell growth followed a Monod growth kinetic model (31, 48). Under a dual nitrogen and phosphorous gradient, a multiplicative model fitted to the growth data showed the colimitation by two nutrients (Supplementary Information). The experimental data presented here inspired us to ask whether there is a general growth kinetic model describing algal cell growth under complex physical and nutrient environments.

**Colimitation of algal growth by light and nitrogen: theoretical modeling**

We propose a general colimitation growth model for algal cells subjected to both physical and chemical parameters. We hypothesized that algal growth response to light or nitrogen was described by Monod kinetics, and the colimitation of algal growth by nitrogen and light was described by a multiplicative model:

$$\mu = \mu_{max}\left(\frac{L+L_0}{K_L+L+L_0}\right)\left(\frac{[N]+N_0}{K_N+[N]+N_0}\right). \quad (1)$$

Here, $\mu_{max}$ is the maximum growth rate, $K_L$ is the half-saturation constant of light intensity, and $K_N$ is the half-saturation constant of nitrogen concentration. We note that the response term to nitrogen is a Monod kinetic model, where $N_0$ represents stored nitrogen. The response term to light is also a Monod kinetic model, with a storage term $L_0$. This term is required since we observed residual growth in the absence of light, but no residual growth in the absence of both light and acetate (Fig. S4). Detailed discussion on the effect of acetate on algal growth can be found in the Supplementary Information.

We fitted Eqn. 1 to experimental data under the dual light and nitrogen gradients. The fit included growth rate data obtained with and without acetate. During fitting, $K_L$ and $K_N$ were kept as free fitting parameters. $\mu_{max}$ was fixed to be 2.4$\text{day}^{-1}$, which was the maximum growth rate obtained in the microhabitats previously (31). $N_0$ was set to 0, as the cells were starved in low N concentration media prior to experiments (See Supplementary Information). In the case with acetate, $L_0$ was left as a free parameter. In the case without acetate, $L_0$ was set to 0 because (1) we observed no growth in the absence of both light and acetate and (2) fitting Eqn. 1 to growth data obtained in the absence of acetate gave a best-fit value of $L_0$ compatible with 0 (p-value = 0.18, Supplementary Information). The fitted surface to Eqn. 1 is plotted together with experimental data in Fig. 5A. The best fit parameters were $L_0$ = 50.8μmol·$m^{-2}$·$s^{-1}$, $K_L$ = 57.2μmol·$m^{-2}$·$s^{-1}$, and $K_N$ = 2.8μM, with standard deviations of 7.9μmol·$m^{-2}$·$s^{-1}$, 8.8μmol·$m^{-2}$·$s^{-1}$, and 0.4μM, respectively. Distributions and correlations of the fitted parameters are shown in Fig. S3.

Colimitation of algal growth by light and nitrogen was revealed by the microhabitat dual gradient experiments and subsequent fitting to a general multiplicative growth model. While growth response to a single resource has often been described via Monod kinetics, here, we showed that growth response to a physical and a chemical parameter could be described by the multiplication of the two Monod kinetics terms with respect to each single resource, known as independent colimitation. Multi-resource colimitation on algal growth has been observed in natural waters, and has been categorized into 1) independent colimitation, where two resources were both at low levels and potentially limiting, and 2) dependent colimitation, where two resources were either biochemical substitutions or biochemically dependent (43-46). Different types of multi-resource-controlled algal growth kinetic models were discussed in the comprehensive reviews by Lee et al. and Bekirogullari et al. (57, 58). As for light and extracellular nitrogen, previous studies that used field data and large-scale culture experiments have described them as independent growth-limiting resources, whose influence could be expressed in a multiplicative form (46, 59-64). However, growth data corresponding to the precise light and nitrogen conditions in the algal microenvironment was unavailable. Our microhabitat platform provides results that fill this gap and demonstrate the independent colimitation of light and nitrogen on the growth of *C. reinhardtii*.

Previous estimates of $K_L$ and $K_N$ in macro-scale systems were 81.4-215 5μmol·$m^{-2}$·$s^{-1}$ and 2.2-17mM respectively (57, 58). These values were larger as compared to the half-saturation constants obtained from our microhabitat platform, and varied in a wide range.

The differences could come from the self-shading effect and varying nutrient concentrations in large scale experiments. We also note that the half-saturation constants here depended on the history of the cell (starved or not), the provided light spectrum, and the form of nitrogen. Previous studies using non-starved cells gave $K_L$ and $K_N$ based on single light and single nitrogen gradients in the microhabitat platform to be 1.9μmol·m$^{-2}$·s$^{-1}$ and 1.2μM, respectively (31, 40), which is smaller than those found in this work. In addition, we also compared different forms of models, including a multiplicative form with only $\mu_0$, $K_L$, and $K_N$ as fitted parameters and a law of minimum form of the growth kinetic model (see Supplemental Information). It was seen that the proposed model in Eqn. 1 had better goodness of fit as compared to the other forms.

**Predicting growth rate under various light intensities and N concentrations using the colimitation model**

The general co-limiting model can be used to predict and understand how light and nutrient synergistically control the growth of algal cells. Using the fitted model, a look up map was generated that predicts algal growth rate given the light intensity and the nitrogen concentration (Fig. 5B). In general, growth is suppressed when either N is under 20 uM, or light intensity is under 20 PAR. This look up map could be used to make predictions on algal growth trends under various light and nitrogen conditions. For example, one could predict the effect of environmental drivers on algal growth in freshwater bodies, where light intensity and nitrogen conditions could vary seasonally, as well as geographically. Take two of the Finger Lakes in upstate New York, USA as an example. In the 2020 sampling season, the light intensity at 6m depth and nitrogen concentration at the lake surface in Cayuga Lake (South Shelf Site) were 0.1PAR and 93μM respectively, while the values for Hemlock Lake (Mid Site) were 100PAR and 17μM (data was taken from the Citizens Statewide Lake Assessment Program reports, conversion from clarity measurements to light intensities could be found in SI). Comparing the growth response to nitrogen concentration at the two different light intensity levels in Fig. 5D, it was shown that nitrogen concentration would be a more important driver for algal growth in Hemlock Lake (100PAR) than in Cayuga Lake (0.1PAR). In addition, comparing the growth response to light intensity at the two different nitrogen concentration levels in Fig. 5E, it was seen that light intensity would be a slightly more important driver in Cayuga Lake (93μM) than in Hemlock Lake (17μM). We note that in natural lakes, the spectrum of light and its attenuation down the water column, the forms of nitrogen that TN involves, and the blooming species are not exactly accounted for in our experiments that led to the model. However, this example shows the potential use of the experimental method and the model to make relevant predictions to some level of generalization.

## Conclusions and future perspectives

Modeling algal growth under the influence of various physical and chemical factors in their microenvironment is important for understanding the ecology and evolution of phytoplankton-related aquatic microbial communities, controlling harmful algal blooms, and improving biofuel production. Here, we implemented a microfluidic platform with precisely controlled light and chemical gradients to study the light and nitrogen-controlled growth of *C. reinhardtii*, then used the experimentally measured data to propose and validate a general growth kinetics model. We found that algal growth response to environmental parameters was described by an independent multiplicative Monod kinetic model. Interestingly, we found that the contributing term to the multiplicative model for both a physical and a chemical parameter were similar, because they can both be described by Monod kinetics. In our experiment, the multiplicative model consists of the Monod growth kinetics due to light multiplied by the Monod growth kinetics due to nitrogen. We hypothesize that more physical and chemical parameters can be included in this general growth model by simply multiplying the Monod kinetics term of each single source together, assuming the resources are independently colimiting. Future experiments will be needed to develop a true general growth model in a complex environment.

In this work, we demonstrated that a microfluidic platform enabled studies of algal growth under a well-defined physical and chemical environment, and that the results were ideal for data driven theoretical modeling. In addition, dynamic information can be obtained through this platform's real time imaging capabilities. This platform can be easily adapted to studies of photosynthetic microbes under two chemical gradients (e.g., nitrogen and phosphorous) alone or with a light gradient. We note that the generated light and ammonium gradients were kept at low levels for the investigation of colimitation effects of the two factors. The range can be easily extended by varying the lamp power output and the concentration of nitrogen in the medium perfused from the side channels. Our platform is also amenable to studies of other cellular behavior such as competition of microbiomes or cell motility. While this small-scale microfluidic device provided a way to precisely define a complex microenvironment for cells, the downside is that there was a small number of cells in each habitat. This was reflected in the variability of the measurements. A future modeling improvement can include cell number variability into the formulation.

A multiplicative growth model of light and nitrogen was used to describe the independent colimitation effect of light and nitrogen on algal growth. The model predictions were used to estimate which environmental driver, nutrient or light, would be the limiting growth factor in two Finger Lakes with different clarities (thus light condition) and different total nitrogen concentrations. The model treated the presence of acetate as contributing to an equivalent baseline light intensity. For independent environmental parameters, the multiplicative model could be further simplified as follows:

$$\mu = \mu_{max} \prod_{i=1}^{n} \left( \frac{E_i + E_{i_o}}{K_{E_i} + E_i + E_{i_o}} \right). \qquad (2)$$

Here, $E_i$ stands for the i-th factor of the $n$ number of independent colimiting factors such as light, nitrogen, phosphorous, and $CO_2$ (46, 65, 66). $K_{E_i}$ is the half saturation constant for the i-th environmental factor, and $E_{i_o}$ is the storage term. $\mu_{max}$ is the maximum growth rate at saturation. Beyond freshwater bodies, mathematical models of phytoplankton growth are being incorporated in global models of ocean circulation and biogeochemistry (67), to better predict the magnitude and spatial intensity of the carbon pump concentration and elemental ratios in the ocean. We anticipate that a better understanding of phytoplankton growth in nutrients and light gradients will provide the foundation for more accurate predictions of the global biogeochemical processes to which phytoplankton contributes.

In response to nitrogen stress, along with the reduction in photosynthesis, algae such as *C. reinhardtii* modify their metabolism to accumulate high amounts of storage molecules including triacylglycerol (TAG), which makes them a promising candidate for biofuel production (7, 9, 49, 68-71). While nitrogen deprivation studies have mainly compared milli molar ammonium to the nitrogen-deprived state, here, we found that an increase of nitrogen concentration in the micro molar range could clearly affect cells utilization of light for growth. This information could possibly be used to search for a nitrogen condition with high TAG accumulation as well as a reasonable growth rate to optimize biofuel production. In addition, optimizing biofuel production requires further investigation of the effect of acetate, where experiments could be performed to figure out $L_0$ proposed in Eqn. 1 as a function of acetate concentration.

## Materials and Methods

### Cell culture and preparation

Wild type *C. reinhardtii* strain CC-125 was obtained from the Stern Laboratory at the Boyce Thompson Institute of Plant Research on the Ithaca campus of Cornell University. Cells were maintained in 10%TAP (Tris Acetate Phosphate) medium, which has one tenth of the acetate concentration as compared to the standard TAP medium. The 10%TAP medium was composed of 2mM Tris, 1.7mM Acetate, 0.68 mM $K_2HPO_4$, 0.45mM $KH_2PO_4$, 7.5 mM $NH_4Cl$, and other salts including 0.34 mM $CaCl_2$), and prepared using an established protocol (72) with trace metal elements concentrations as described in Hunter et al. (73) A 5mL cell culture was maintained in 15mL glass tubes in a temperature-controlled incubator at 25°C without shaking (New Brunswick Innova 44, Eppendorf) under a continuous illumination of 20 $\mu mol \cdot m^{-2} \cdot s^{-1}$ using LED light (4000K, Commercial Electric). For experiments in Tris-Minimal medium (TM, TAP without acetate), cells were maintained in a similar manner with TM instead of 10%TAP.

To make medium with low N concentrations, the $NH_4Cl$ in the Beijerinck's solution in the 10%TAP or TM recipe was replaced by NaCl. 750mM $NH_4Cl$ stock solution was made and added to the medium to achieve final N concentrations in the micro molar range. The medium with no $NH_4Cl$ stock solution added was referred to as 10%TAP-N or TM-N, which had the lowest nitrogen concentrations of 5.3uM due to the ammonium salt in the trace element solution in the recipe. $NH_4^+$ was the only nitrogen source in the medium, thus its concentration was taken as the nitrogen concentration.

To set up experiment in the microfluidic platform, cells were prepared via three steps: (i) a 20-day old maintenance culture was diluted 10X into 5mL of fresh 10%TAP medium; (ii) the newly transferred culture was kept for 5 days before being centrifuged and washed twice, and then diluted 5X into 5mL of 10%TAP-N medium; (iii) two days after limiting cells of nitrogen, the culture was concentrated 10X to a final density of about $10^6$ cells per mL and used to seed the microhabitats. All the glassware for the nitrogen-limitation experiment was washed with 10% HCl to remove possible nutrient residuals. Experiments in TM were conducted following the same procedures with TM and TM-N.

### Microfluidic device design, fabrication, and assembly

The microfluidic device design consists of an array of 8 x 8 microhabitats (100μm x 100μm x 100μm each) surrounded by two sets of side channels (400μm W x 200μm H) imprinted in agarose gel, and was proved in previous work to provide stable and well-controlled single and dual chemical gradients via molecular diffusion (40).

A two-layer SU8 negative photoresist photolithography process was used to fabricate the silicon master with the desired pattern. The pattern was transferred to a 1mm film of agarose gel by pouring 3% dissolved agarose in 1X PBS (Phosphate-Buffered Saline) onto the silicon master and letting it cure under room temperature. The gel was then soaked in 10%TAP-N medium overnight.

To assemble the device, 200uL cells ($1*10^6$ cells/mL) were seeded into the agarose gel with patterned microhabitats. The gel was then sandwiched between a glass slide and a Plexi glass manifold, and clamped using a metal frame and screws to seal the microhabitats and side channels. After initial seeding, there were usually 1 to 6 cells in each microhabitat. Empty microhabitats due to randomness of the seeding were omitted in the data analysis, as well as microhabitats in which cells that had been trapped in the surrounding gel during seeding started to grow into the microhabitat throughout the experiment.

**Experimental setup and gradient generations**

The microfluidic device was kept on the microscope stage throughout the experiment (7 days). One set of the side channels (top and bottom) was used for medium perfusion and nitrogen gradient generation, while the other set had ends plugged to prevent evaporation. A constant flow rate at 0.7 µL/min through the side channels was maintained by a syringe pump (KDS230, KD Scientific, Holliston, MA) and two 10 mL syringes (Exelint International Co., Redondo Beach, CA). For the nitrogen gradient experiment, the syringe connected to the top channel was filled with 10%TAP with 35.3µM N, while the syringe connected to the bottom channel was filled with 10%TAP-N. For the lowest nitrogen (5.3µM) condition, the source and sink channels were both perfused with 10%TAP-N, with habitats either under a light intensity gradient, or under dark (0 PAR).

The transmitted light path of an inverted microscope (Olympus IX81, Center Valley, CA) was modified to generate a light gradient as described previously(48). Briefly, a half-moon mask was inserted into the light path, perpendicular to the light beam from the bright field halogen lamp (Olympus U-LH100L-3), which resulted in a light gradient with a light-dark transition region near the middle of the sample plane. The microscope room temperature was controlled at 25°C.

**Imaging and data analysis**

Microscopic images were taken by an EMCCD camera (ImagEM X2 EM-CCD camera, Hamamatsu Photonics K.K.). For fluorescence imaging of *C. reinhardtii* cells, a fluorescence lamp (X-Cite 120PC Q, Excelitas Technologies Corp.), a 488/10 nm single bandpass excitation filter (Semrock, Rochester, NY), and a 440/521/607/700 quad-bandpass emission filter (Semrock, Rochester, NY) were used. During the 7-day experiment period, fluorescence images were taken every 4 hours with 50ms exposure using the cellSens imaging software (Olympus Life Science).

Fluorescence intensities of microhabitats were measured using the software ImageJ and the background was subtracted for use as a measure proportional to cell number, as validated by cell counts in bright field images. The growth rates were obtained by fitting n consecutive data points of $\ln\frac{N}{N0}$ versus time in the growth curve of each microhabitat to a linear function, and finding the maximum after 1.3 days (Fig. S2). For our data analysis, n=9 was chosen because (1) it reduces the noise in measurements, as the calculated growth rates versus time curves flattened out as n increased from 5 (Fig. S2A), and (2) further increasing n until 25 did not overly change the maximum growth rate values obtained, indicating only marginal benefit in noise reducing for n values larger than 9 (Fig. S2B,C).

Growth kinetics model fitting was done using a bootstrapping method on three replicate experimental datasets. For each bootstrapping run, a growth rate matrix at the tested combinations of light and nitrogen conditions was constructed by randomly sampling from the three datasets with a weight given by the reciprocal of the squared standard error from 4 points around each maximum growth rate. The constructed growth rate matrix was then used to fit the multiplicative model (equation 1) using nonlinear least square fit function in MATLAB, and the fitting parameters ($L_0$, $K_L$, and $K_N$) were recorded. This process was repeated 1000 times, which yield a distribution for each parameter (Fig. S3). The estimated parameters were calculated by taking the average of the 1000 values.

## Acknowledgments

MW and FL would like to thank Beth Ahner, Mohammad Yazdani, Steve Winans and Nelson Hairston for many helpful discussions throughout this project. This work is supported by the USDA National Institute of Food and Agriculture, AFRI project [2016-08830], the Academic Venture Fund from the Cornell Atkinson Center for Sustainability, and The New York State Hatch fund. This work was performed in part at the Cornell NanoScale Facility, a member of the National Nanotechnology Coordinated Infrastructure (NNCI), which is supported by the National Science Foundation (Grant NNCI-2025233). FL would like to thank CNF staff member Tom Pennell for helping with the microfabrication processes. AG acknowledges support from NIH NIGMS grant 1R35GM147493-01.

## References

1. J. Huisman *et al.*, Cyanobacterial blooms. *Nat Rev Microbiol* **16**, 471-483 (2018).
2. H. W. Paerl, T. G. Otten, R. Kudela, Mitigating the Expansion of Harmful Algal Blooms Across the Freshwater-to-Marine Continuum. *Environ Sci Technol* **52**, 5519-5529 (2018).
3. M. L. Wells *et al.*, Harmful algal blooms and climate change: Learning from the past and present to forecast the future. *Harmful Algae* **49**, 68-93 (2015).
4. B. Qin *et al.*, Extreme Climate Anomalies Enhancing Cyanobacterial Blooms in Eutrophic Lake Taihu, China. *Water Resources Research* **57**, e2020WR029371 (2021).
5. M. L. Wells *et al.*, Future HAB science: Directions and challenges in a changing climate. *Harmful Algae* **91**, 101632 (2020).
6. H. W. Paerl, M. A. Barnard, Mitigating the global expansion of harmful cyanobacterial blooms: Moving targets in a human- and climatically-altered world. *Harmful Algae* **96**, 101845 (2020).
7. Q. Hu *et al.*, Microalgal triacylglycerols as feedstocks for biofuel production: perspectives and advances. *The Plant Journal* **54**, 621-639 (2008).
8. C. S. Jones, S. P. Mayfieldt, Algae biofuels: versatility for the future of bioenergy. *Curr Opin Biotech* **23**, 346-351 (2012).
9. R. H. Wijffels, M. J. Barbosa, An Outlook on Microalgal Biofuels. *Science* **329**, 796-799 (2010).
10. C. Enzing, M. Ploeg, M. Barbosa, L. Sijtsma, Microalgae-based products for the food and feed sector: an outlook for Europe. *JRC Scientific and policy reports*, 19-37 (2014).
11. S. Buono, A. L. Langellotti, A. Martello, F. Rinna, V. Fogliano, Functional ingredients from microalgae. *Food & Function* **5**, 1669-1685 (2014).
12. Y. Torres-Tiji, F. J. Fields, S. P. Mayfield, Microalgae as a future food source. *Biotechnology Advances* **41**, 107536 (2020).
13. S. F. Gilbert, T. C. G. Bosch, C. Ledón-Rettig, Eco-Evo-Devo: developmental symbiosis and developmental plasticity as evolutionary agents. *Nature Reviews Genetics* **16**, 611-622 (2015).
14. S. Skúlason *et al.*, A way forward with eco evo devo: an extended theory of resource polymorphism with postglacial fishes as model systems. *Biological Reviews* **94**, 1786-1808 (2019).
15. N. Rivera-Yoshida, A. Hernandez-Teran, A. E. Escalante, M. Benitez, Laboratory biases hinder Eco-Evo-Devo integration: Hints from the microbial world. *Journal of Experimental Zoology Part B: Molecular and Developmental Evolution* **334**, 14-24 (2020).
16. L. Feng *et al.*, An eco-evo-devo genetic network model of stress response. *Horticulture Research* **9**, uhac135 (2022).
17. A. Watkins, Development and microbiology. *Biology & Philosophy* **36**, 34 (2021).
18. W. E. Bradshaw, P. A. Zani, C. M. Holzapfel, Adaptation to temperate climates. *Evolution* **58**, 1748-1762 (2004).
19. C. K. Ghalambor, J. K. McKay, S. P. Carroll, D. N. Reznick, Adaptive versus non-adaptive phenotypic plasticity and the potential for contemporary adaptation in new environments. *Functional Ecology* **21**, 394-407 (2007).
20. A. Charmantier *et al.*, Adaptive phenotypic plasticity in response to climate change in a wild bird population. *Science* **320**, 800-803 (2008).

21. E. Jablonka, M. J. Lamb, *Evolution in Four Dimensions: Genetic, Epigenetic, Behavioral, and Symbolic Variation in the History of Life* (MIT Press, 2014), pp. 1-563.
22. M. McFall-Ngai *et al.*, Animals in a bacterial world, a new imperative for the life sciences. *Proc Natl Acad Sci U S A* **110**, 3229-3236 (2013).
23. R. Guerrero, L. Margulis, M. Berlanga, Symbiogenesis: the holobiont as a unit of evolution. *Int Microbiol* **16**, 133-143 (2013).
24. L. A. Richardson, Evolving as a holobiont. *PLOS Biology* **15**, e2002168 (2017).
25. D. Skillings, Holobionts and the ecology of organisms: Multi-species communities or integrated individuals? *Biology & Philosophy* **31**, 875-892 (2016).
26. F. J. Odling-Smee, K. N. Laland, M. W. Feldman, *Niche Construction: The Neglected Process in Evolution (MPB-37)* (Princeton University Press, 2003).
27. J. B. Saltz, S. V. Nuzhdin, Genetic variation in niche construction: implications for development and evolutionary genetics. *Trends Ecol Evol* **29**, 8-14 (2014).
28. B. J. Callahan, T. Fukami, D. S. Fisher, Rapid evolution of adaptive niche construction in experimental microbial populations. *Evolution* **68**, 3307-3316 (2014).
29. S. Pajares, R. Ramos, Processes and Microorganisms Involved in the Marine Nitrogen Cycle: Knowledge and Gaps. *Front Mar Sci* **6** (2019).
30. H. S. Kim, A. R. Guzman, H. R. Thapa, T. P. Devarenne, A. Han, A droplet microfluidics platform for rapid microalgal growth and oil production analysis. *Biotechnol Bioeng* **113**, 1691-1701 (2016).
31. B. J. Kim *et al.*, An array microhabitat system for high throughput studies of microalgal growth under controlled nutrient gradients. *Lab Chip* **15**, 3687-3694 (2015).
32. S. Bae, C. W. Kim, J. S. Choi, J. W. Yang, T. S. Seo, An integrated microfluidic device for the high-throughput screening of microalgal cell culture conditions that induce high growth rate and lipid content. *Anal Bioanal Chem* **405**, 9365-9374 (2013).
33. C. C. Yang, R. C. Wen, C. R. Shen, D. J. Yao, Using a Microfluidic Gradient Generator to Characterize BG-11 Medium for the Growth of Cyanobacteria Synechococcus elongatus PCC7942. *Micromachines-Basel* **6**, 1755-1767 (2015).
34. H. S. Kim, T. L. Weiss, H. R. Thapa, T. P. Devarenne, A. Han, A microfluidic photobioreactor array demonstrating high-throughput screening for microalgal oil production. *Lab Chip* **14**, 1415-1425 (2014).
35. P. J. Graham, J. Riordon, D. Sinton, Microalgae on display: a microfluidic pixel-based irradiance assay for photosynthetic growth. *Lab Chip* **15**, 3116-3124 (2015).
36. D. Kilinc *et al.*, A microfluidic dual gradient generator for conducting cell-based drug combination assays. *Integrative Biology* **8**, 39-49 (2015).
37. R. Guo, C.-G. Yang, Z.-R. Xu, A superposable double-gradient droplet array chip for preparation of PEGDA microspheres containing concentration-gradient drugs. *Microfluidics and Nanofluidics* **21**, 157 (2017).
38. X. J. Liu *et al.*, Microfluidic liquid-air dual-gradient chip for synergic effect bio-evaluation of air pollutant. *Talanta* **182**, 202-209 (2018).
39. B. Nguyen, P. J. Graham, C. M. Rochman, D. Sinton, A Platform for High-Throughput Assessments of Environmental Multistressors. *Adv Sci* **5**, 1700677 (2018).
40. F. Liu, M. Yazdani, B. A. Ahner, M. Wu, An array microhabitat device with dual gradients revealed synergistic roles of nitrogen and phosphorous in the growth of microalgae. *Lab Chip* **20**, 798-805 (2020).

41. M. G. Saad *et al.*, High-Throughput Screening of Chlorella Vulgaris Growth Kinetics inside a Droplet-Based Microfluidic Device under Irradiance and Nitrate Stress Conditions. *Biomolecules* **9**, 276 (2019).
42. B. Nguyen, P. J. Graham, D. Sinton, Dual gradients of light intensity and nutrient concentration for full-factorial mapping of photosynthetic productivity. *Lab Chip* **16**, 2785-2790 (2016).
43. K. R. Arrigo, Marine microorganisms and global nutrient cycles. *Nature* **437**, 349-355 (2005).
44. E. T. Buitenhuis, K. R. Timmermans, H. J. W. de Baar, Zinc-bicarbonate colimitation of Emiliania huxleyi. *Limnol Oceanogr* **48**, 1575-1582 (2003).
45. E. Spijkerman, F. de Castro, U. Gaedke, Independent Colimitation for Carbon Dioxide and Inorganic Phosphorus. *Plos One* **6**, e28219 (2011).
46. M. A. Saito, T. J. Goepfert, J. T. Ritt, Some thoughts on the concept of colimitation: Three definitions and the importance of bioavailability. *Limnol Oceanogr* **53**, 276-290 (2008).
47. E. Lee, M. Jalalizadeh, Q. Zhang, Growth kinetic models for microalgae cultivation: A review. *Algal Res* **12**, 497-512 (2015).
48. F. Liu, L. Gaul, F. Shu, D. Vitenson, M. Wu, Microscope-based light gradient generation for quantitative growth studies of photosynthetic micro-organisms. *Lab Chip* **22**, 3138-3146 (2022).
49. M. T. Juergens *et al.*, The Regulation of Photosynthetic Structure and Function during Nitrogen Deprivation in Chlamydomonas reinhardtii. *Plant Physiology* **167**, 558-573 (2014).
50. S. Schmollinger *et al.*, Nitrogen-Sparing Mechanisms in Chlamydomonas Affect the Transcriptome, the Proteome, and Photosynthetic Metabolism. *The Plant Cell* **26**, 1410-1435 (2014).
51. N. Wase, P. N. Black, B. A. Stanley, C. C. DiRusso, Integrated Quantitative Analysis of Nitrogen Stress Response in Chlamydomonas reinhardtii Using Metabolite and Protein Profiling. *Journal of Proteome Research* **13**, 1373-1396 (2014).
52. J.-J. Park *et al.*, The response of Chlamydomonas reinhardtii to nitrogen deprivation: a systems biology analysis. *The Plant Journal* **81**, 611-624 (2015).
53. T. Mettler *et al.*, Systems Analysis of the Response of Photosynthesis, Metabolism, and Growth to an Increase in Irradiance in the Photosynthetic Model Organism Chlamydomonas reinhardtii. *The Plant Cell* **26**, 2310-2350 (2014).
54. S. Imam *et al.*, A refined genome-scale reconstruction of Chlamydomonas metabolism provides a platform for systems-level analyses. *The Plant Journal* **84**, 1239-1256 (2015).
55. C. Shene, J. A. Asenjo, Y. Chisti, Metabolic modelling and simulation of the light and dark metabolism of Chlamydomonas reinhardtii. *The Plant Journal* **96**, 1076-1088 (2018).
56. A. J. Márquez, F. Galvan, J. M. Vega, Utilization of Ammonium by Mutant and Wild Types of Chlamydomonas reinhardii. Studies of the Glutamate Synthase Activities. *Journal of Plant Physiology* **124**, 95-102 (1986).
57. E. Lee, M. Jalalizadeh, Q. Zhang, Growth kinetic models for microalgae cultivation: A review. *Algal Research* **12**, 497-512 (2015).
58. M. Bekirogullari, G. M. Figueroa-Torres, J. K. Pittman, C. Theodoropoulos, Models of microalgal cultivation for added-value products - A review. *Biotechnology Advances* **44**, 107609 (2020).

59. H. Haario, L. Kalachev, M. Laine, Reduced Models of Algae Growth. *Bulletin of Mathematical Biology* **71**, 1626-1648 (2009).
60. A. Yang, Modeling and Evaluation of CO2 Supply and Utilization in Algal Ponds. *Industrial & Engineering Chemistry Research* **50**, 11181-11192 (2011).
61. B. Ketheesan, N. Nirmalakhandan, Modeling microalgal growth in an airlift-driven raceway reactor. *Bioresource Technol* **136**, 689-696 (2013).
62. Y.-H. Wu *et al.*, An integrated microalgal growth model and its application to optimize the biomass production of Scenedesmus sp. LX1 in open pond under the nutrient level of domestic secondary effluent. *Bioresource Technol* **144**, 445-451 (2013).
63. M. Bekirogullari, I. S. Fragkopoulos, J. K. Pittman, C. Theodoropoulos, Production of lipid-based fuels and chemicals from microalgae: An integrated experimental and model-based optimization study. *Algal Research* **23**, 78-87 (2017).
64. A. Franz, F. Lehr, C. Posten, G. Schaub, Modeling microalgae cultivation productivities in different geographic locations – estimation method for idealized photobioreactors. *Biotechnology Journal* **7**, 546-557 (2012).
65. C. R. Benitez-Nelson, The biogeochemical cycling of phosphorus in marine systems. *Earth-Science Reviews* **51**, 109-135 (2000).
66. M. Hein, K. Sand-Jensen, CO2 increases oceanic primary production. *Nature* **388**, 526-527 (1997).
67. K. Inomura, C. Deutsch, O. Jahn, S. Dutkiewicz, M. J. Follows, Global patterns in marine organic matter stoichiometry driven by phytoplankton ecophysiology. *Nature Geoscience* **15**, 1034-1040 (2022).
68. N. R. Boyle *et al.*, Three Acyltransferases and Nitrogen-responsive Regulator Are Implicated in Nitrogen Starvation-induced Triacylglycerol Accumulation in Chlamydomonas. *Journal of Biological Chemistry* **287**, 15811-15825 (2012).
69. S. S. Merchant, J. Kropat, B. Liu, J. Shaw, J. Warakanont, TAG, You're it! Chlamydomonas as a reference organism for understanding algal triacylglycerol accumulation. *Curr Opin Biotech* **23**, 352-363 (2012).
70. B. Liu, C. Benning, Lipid metabolism in microalgae distinguishes itself. *Curr Opin Biotech* **24**, 300-309 (2013).
71. N. S. Shifrin, S. W. Chisholm, PHYTOPLANKTON LIPIDS: INTERSPECIFIC DIFFERENCES AND EFFECTS OF NITRATE, SILICATE AND LIGHT-DARK CYCLES1. *J Phycol* **17**, 374-384 (1981).
72. R. K. Togasaki, K. Brunke, M. Kitayama, O. M. Griffith, "Isolation of Intact Chloroplasts from Chlamydomonas Reinhardtii with Beckman Centrifugal Elutriation System" in Progress in Photosynthesis Research: Volume 3 Proceedings of the VIIth International Congress on Photosynthesis Providence, Rhode Island, USA, August 10–15, 1986, J. Biggins, Ed. (Springer Netherlands, Dordrecht, 1987), 10.1007/978-94-017-0516-5_106, pp. 499-502.
73. S. H. Hutner, Organic Growth Essentials of the Aerobic Nonsulfur Photosynthetic Bacteria. *J Bacteriol* **52**, 213-221 (1946).

**Figures**

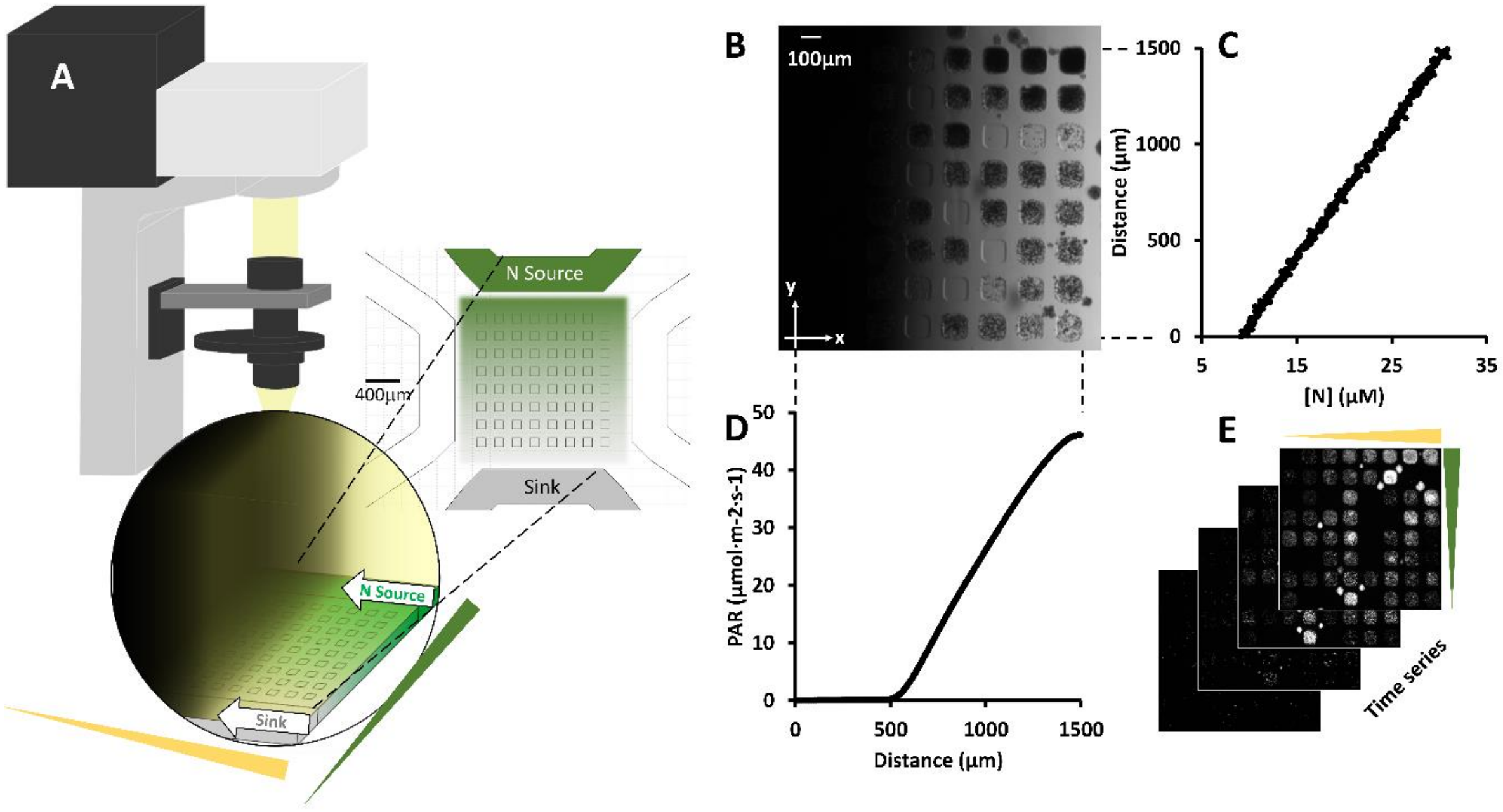

**Figure 1. Generation of a dual light and nitrogen gradient for algal cells. (A)** Schematics of the experimental setup. The transmitted illumination light path of an inverted microscope was modified to provide a light gradient. A hydrogel-based array microhabitat platform was used to provide a nitrogen concentration gradient for algal cells. Here, media with a constant nitrogen concentration and buffer flow through the source and sink channels, respectively, and a nitrogen concentration gradient was established in the array microhabitat via diffusion. The blank side channels were filled with blank media and plugged. **(B)** A bright field image of the 8 x 8 microhabitat array on day 7. The light intensity gradient was in the x-direction and the nitrogen concentration gradient was in the y-direction. *Chlamydomonas reinhardtii* cells were seeded in the array microhabitats on day 0. **(C)** Characterization of the nitrogen gradient in the microhabitat array. Experimental nitrogen concentration ([N]) gradient along the y direction across the microhabitat array. The linear gradient was computed using images taken when the fluorescent dye and blank buffer flow through the source and sink channels respectively. Mean [N] at the horizontal centerline of each row of microhabitats was used as the nitrogen concentration for the row. **(D)** Light intensity gradient in the x-direction as measured by grayscale values of bright field images of the microhabitat array. The PAR (µmol·m$^{-2}$·s$^{-1}$) values were converted from grayscale values using a linear relationship obtained in a previous work(48). Mean light intensity at the vertical centerline of each column of microhabitats was used as the light intensity for the column. **(E)** Time lapse fluorescence images of the microhabitats under dual light and nitrogen gradients. Images at 0, 14, 28, 42 hours are shown here.

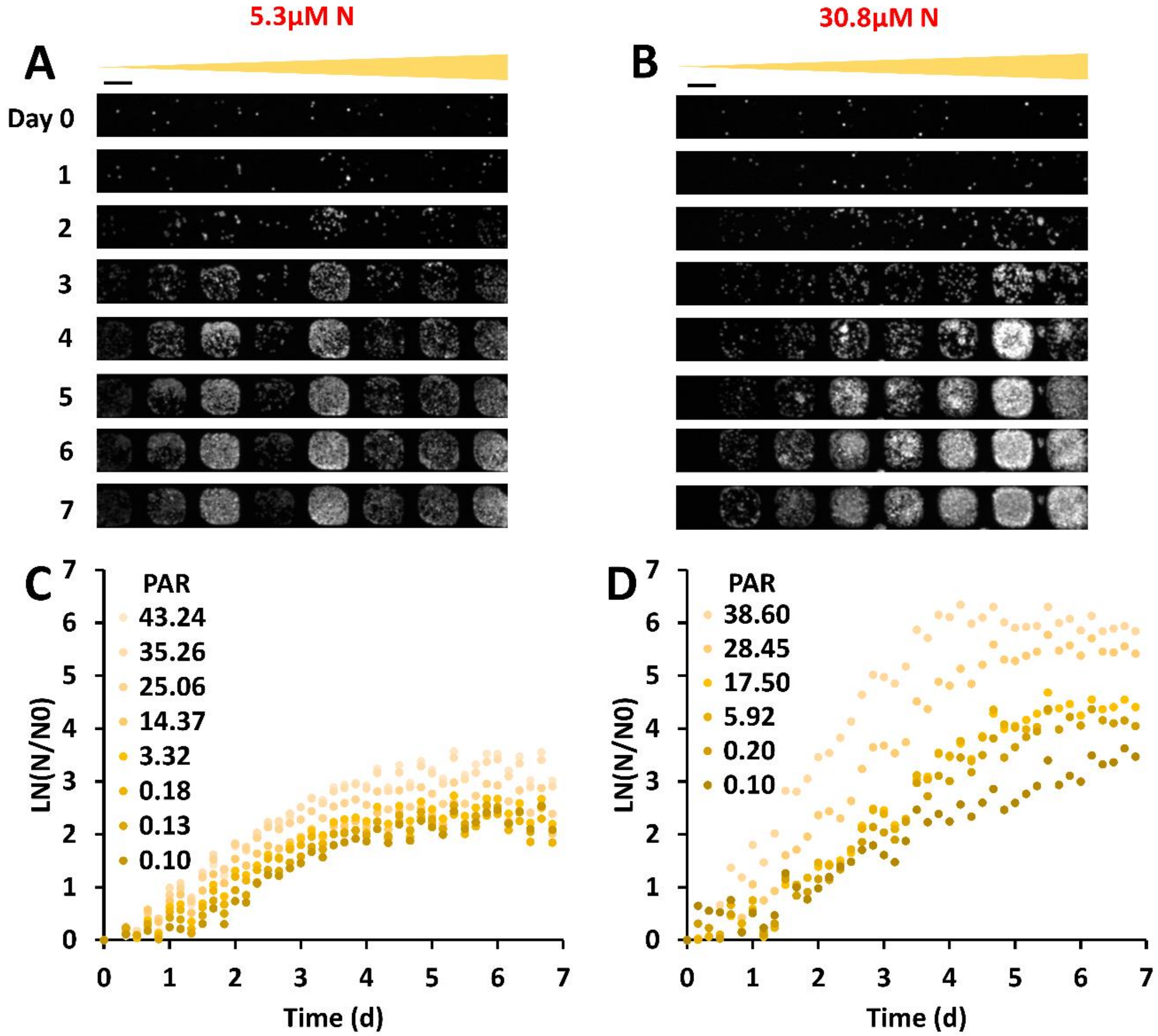

**Figure 2. Nitrogen enhanced the algal growth response sensitivity to light. (A,B)** Time sequence of fluorescence images of one row of algal cells under a light intensity gradient with low (5.3μM) (A) and high (30.8μM) (B) nitrogen concentration. Contrast was adjusted to show details: Maximum grayscale value was taken as 15000 for day 0-4 images, and 35000 for day 5-7 images. The scale bar represents 100μm. **(C,D)** Growth curves of algal cells in the microhabitats at various levels of light intensity for low (C) and high (D) nitrogen concentrations respectively. Growth was calculated by taking the natural log of cell number N divided by the initial cell number $N_0$, as measured by the fluorescence intensity. We note that initial cell seeding was a random process that could result in variations in cells density. This heterogeneity was addressed by normalizing N by the initial cell number, $N_0$. Colors of the dots represent light intensities. Photosynthetically Active Radiation (PAR) is expressed in units of $\mu mol \cdot m^{-2} \cdot s^{-1}$. Results shown are from one of the three replicates.

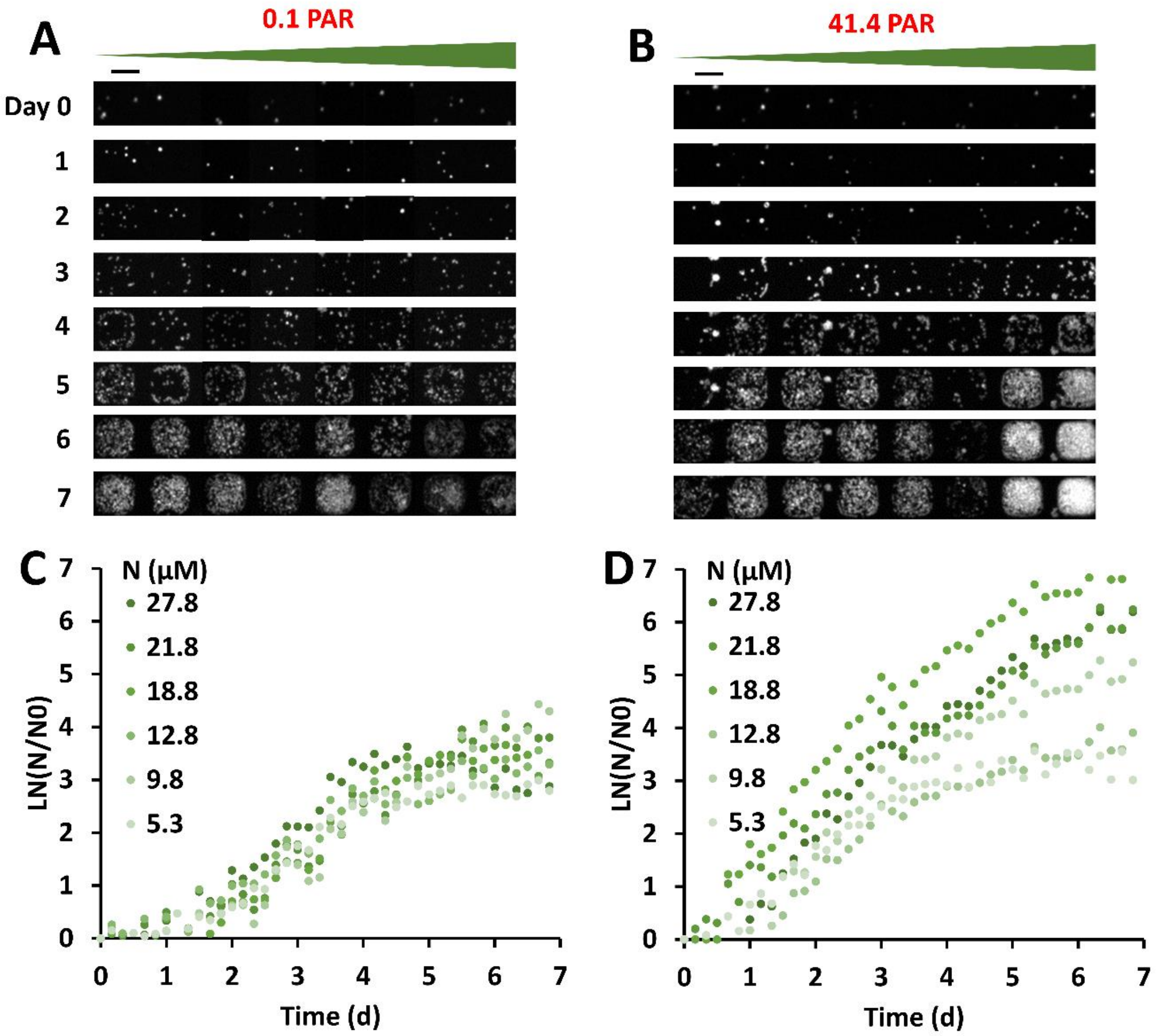

**Figure 3. Light enhanced the algal growth response sensitivity to nitrogen. (A,B)** Time sequence of fluorescence images of algal cells in one column of microhabitats (rotated 90 degrees into a row) under a nitrogen concentration gradient with low (0.1 PAR, A) and high (41.1 PAR,B) light intensities. Contrast was adjusted to show details: Maximum grayscale value was taken as 15000 for day 0-4 images, and 35000 for day 5-7 images. The scale bar represents 100µm. **(C,D)** Growth curves of algal cells at various levels of nitrogen concentration in the microhabitats for low (C) and high (D) light intensities. Growth was calculated by taking the natural log of cell number N divided by the initial cell number $N_0$, as measured by fluorescence intensity. Colors of the dots represent nitrogen concentration. Photosynthetically Active Radiation (PAR) is expressed in units of $\mu mol \cdot m^{-2} \cdot s^{-1}$. Results shown are from one of the three replicates.

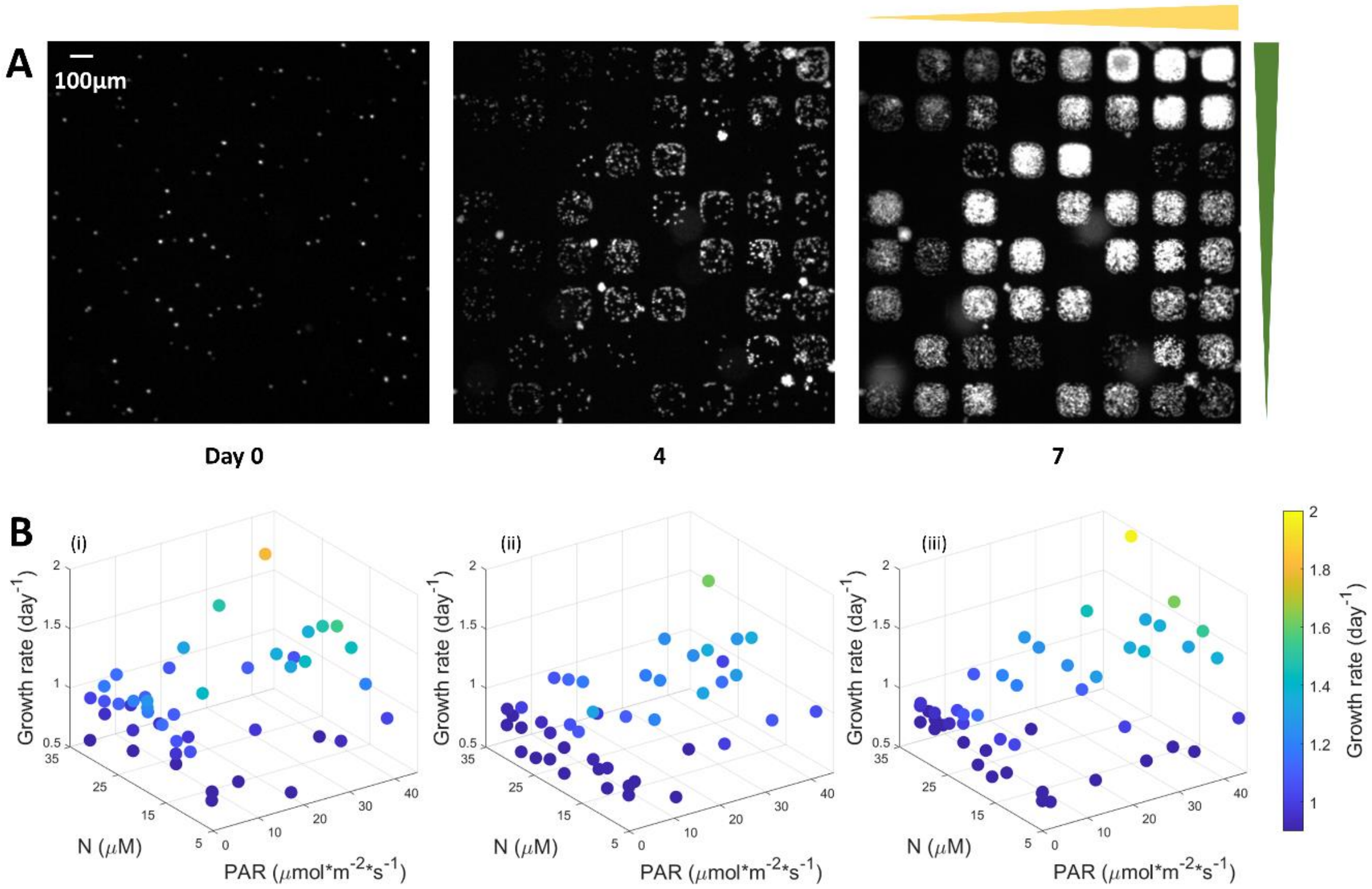

**Figure 4. Light and nitrogen synergistically enhanced algal growth. (A)** Fluorescence images of algal cells growing under dual light and nitrogen gradients (as described in Figure 1) on day 0, 4 and 7. The nitrogen gradient was 1.3µM per 100µm and the light intensity gradient is 2 $\mu mol\cdot m^{-2}\cdot s^{-1}$ per 100µm. Contrast was adjusted for illustration purposes: the intensity (0-255) corresponds to (0-10000) of the grayscale of original images for day 0 and 4, and (0-20000) for day 7. **(B)** 3D plots of cell growth rates in single microhabitats. Data were combined using results from three replicated experiments. The color of dots represents growth rate as shown in the color bar.

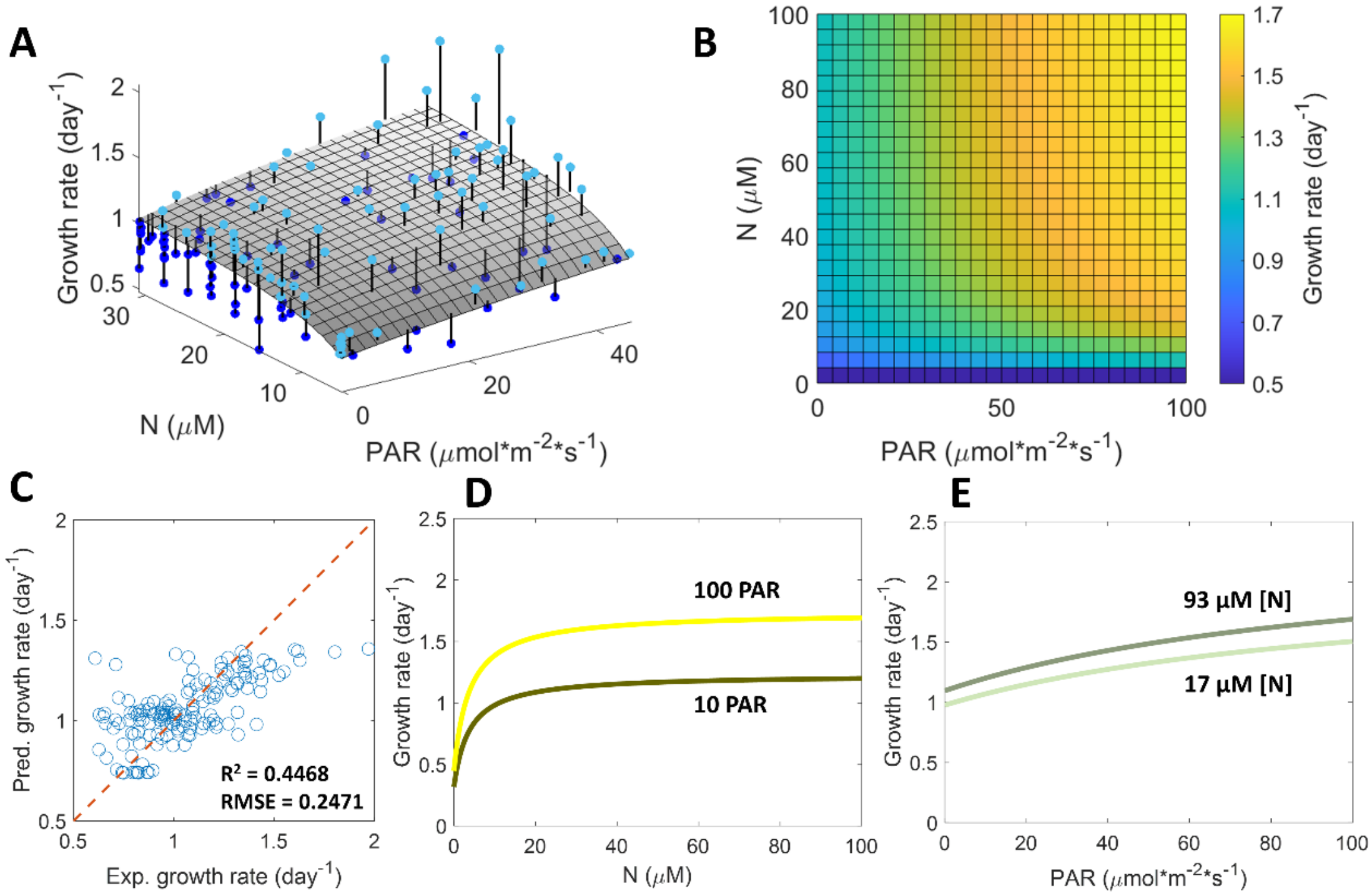

**Figure 5. Light and nitrogen co-limit algal cell growth, revealed by a general multiplicative growth model. (A)** 3D plot of the growth rate as a function of nitrogen concentration and light intensity. The dots are experimental data from three replicates, and the surface is a fit to the proposed general multiplicative model. Experimental dots are connected to the fitted surface by vertical lines for visualization purpose. Dots above or on the surface are plotted in light blue, while those below are in dark blue. The fitted parameters were $L_0$ = 50.8 μmol·m$^{-2}$·s$^{-1}$, $K_L$ = 57.2 μmol·m$^{-2}$·s$^{-1}$, and $K_N$ = 2.8 μM. **(B)** A lookup map showing the predicted growth rate under wide ranges of light intensities and nitrogen concentrations using the multiplicative model. **(C)** Scatter plots of predicted growth rates from the multiplicative model versus experimental growth rates. **(D)** Predicted algal growth rate as a function of nitrogen concentration under low (10 PAR) and high (100 PAR) light intensities. (**E**) Predicted growth rate as a function of light intensity under low (17 μM) and high (93 μM) nitrogen concentrations.

**Supporting Information for**

# Colimitation of light and nitrogen on algal growth revealed by an array microhabitat platform

Fangchen Liu[1], Larissa Gaul[1], Andrea Giometto[2]* and Mingming Wu[1]*
[1]Department of Biological and Environmental Engineering, Cornell University, Ithaca, NY, USA.
[2]School of Civil and Environmental Engineering, Cornell University, Ithaca, NY, USA

*Mingming Wu
*Email: mw272@cornell.edu
*Andrea Giometto
*Email: giometto@cornell.edu

**This PDF file includes:**

Supporting text
Figures S1 to S5
Legends for Movie S1
SI References

**Other supporting materials for this manuscript include the following:**

Movies S1

## Supporting Information Text

### A general multiplicative model considering acetate and nutrient storage

Acetate is an organic carbon source that can provide energy to *C. reinhardtii* cells, which likely affect algal growth response to light. We note that the presence of acetate in the natural environment is typically very low. So, in our experiments, we used a lowered acetate concentration as compared to the normal TAP medium. It has been shown that microalgae like *C. reinhardtii* can have different trophic modes depending on the type of carbon and energy sources available (1, 2). Cells are known to perform mixotrophic growth in presence of both acetate and light, where both inorganic and organic carbon could be used as carbon sources, and both organic carbon and light as energy sources. In algal cells, acetate is incorporated into acetyl coenzyme A (acetyl-CoA), which then goes into the mitochondrial tricarboxylic acid (TCA) cycle for respiration and ATP production (energy source) or into the glyoxylate cycle for gluconeogenesis (C source)(3). The presence of acetate in the growth medium was found to stimulate growth, promoting respiration while reducing chlorophyll content and photosynthesis capacity (1, 4). Therefore, as compared to growth in the absence of acetate, cells growing in mixotrophy could have higher growth rate under the same light condition and appear less sensitive to changes in light. In addition, the presence of acetate alone could fulfill the requirement of nitrogen assimilation on both light and $CO_2$ (3, 5). Thus, acetate could be viewed as a substitute for light and carbon, which is likely an independent growth limiting factor from nitrogen.

To look at the impact of organic carbon on the algal growth response to light, we performed experiments in the microhabitats using Tris-Minimal (TM) medium under a light gradient, where no acetate was provided, and compared to cells growing in 10%TAP (Fig. S4, 10%TAP-N and TM-N refer to media with no NH4Cl stock solution added, and the nitrogen concentration was 5.3μM in both conditions). It was found that removal of the organic carbon resulted in no baseline growth at PAR=0 (Fig. S4F), indicating photoautotrophic growth where inorganic carbon ($CO_2$) was the only carbon source, and light the only energy source. Interestingly, cells in TM responded more sensitively to light as compared to those in 10%TAP, as shown in Fig. S4. Therefore, in the growth model, a storage term $L_0$ was added to address the effect of acetate. In the literature, organic carbon has been mostly treated as an independent factor to light either in a multiplicative manner (6-9), where the kinetic terms for the organic carbon and light were multiplied, or in an additive manner(10, 11), where the contributions from phototrophic and heterotrophic growth were added together. Here, based on the comparison between the two conditions in TM and 10%TAP, we proposed to describe the effect of acetate as $L_0$ that addressed the decreased growth sensitivity to light, as well as the elevated growth rate as compared to photoautotrophic growth at the same light intensities (Eqn. 1). These growth responses agreed with the growth and metabolic behavior observed in large scale cultures. For the light gradient experiment in TM-N, Eqn. 1 can be written as:

$$\mu = \mu_{max}' \frac{L+L_0}{K_L+L+L_0}. \quad \text{(S1)}$$

Here, $\mu_{max}'$ is the growth rate at saturated light intensity at 5.3μM nitrogen and was fixed to be 0.55day$^{-1}$. Fitting Eqn. S1 gave $L_0$ = 1.6±1.3μmol·m$^{-2}$·s$^{-1}$ and $K_L$ = 15.4±4.3μmol·m$^{-2}$·s$^{-1}$. The obtained value of $L_0$ was not significantly different from 0 (p-value = 0.18). Therefore, when fitting Eqn.1 to the combined datasets obtained with acetate (in 10%TAP) and without (in TM), $L_0$ was kept as a free positive parameter for the former, while set to 0 for the latter. Note that to figure out $L_0$ as a function of acetate concentration, experiments on a range of acetate conditions would be required.

In addition to 1) light and nitrogen, a physical factor and a chemical factor independently colimiting algal growth, and 2) acetate and light, a chemical resource affecting the effect of physical light, the effect of phosphorous and nitrogen, two chemical resources, on algal growth has been explored in this array microhabitat platform with dual chemical gradients. It was found that phosphorus and nitrogen synergistically promoted cell growth(12). The observed non-zero residual growth at [P],[N]=0μM led to the hypothesis that cells had stored phosphorus and nitrogen available despite the starvation on the two nutrients ahead of the experiments. Therefore, the model we proposed to describe the nitrogen and phosphorous colimited growth is:

$$\mu = \mu_{max} \cdot \frac{[P]+P_0}{K_P+[P]+P_0} \cdot \frac{[N]+N_0}{K_N+[N]+N_0}. \quad \text{(S2)}$$

Here, $P_0$ and $N_0$ can be understood as equivalent external concentrations representing nutrient storage. Fitting equation S2 with fixed $\mu_{max}$ gave $P_0$ = 8.0±2.1μM, $N_0$ = 0.97±0.89μM, $K_P$ = 10.2±2.2μM, and $K_N$ = 0.37±0.35μM. The fitting results were shown in Fig. S5 (A-B). An alternative way to address the observed residual growth is to include a $\mu_0$ instead of $P_0$ and $N_0$ in the model, as shown in the following equation:

$$\mu = \mu_{max} + \mu_{max} \cdot \frac{[P]}{K_P+[P]} \cdot \frac{[N]}{K_N+[N]}. \quad \text{(S3)}$$

However, model as described by Eqn. S2 provided better goodness of fit, shown by the reduced Akaike information criterion value, as compared to Eqn. S3 (Fig. S5C-D).

Taken together, the growth contribution from acetate and nutrient storage discussed above, assuming that (1) light and nitrogen independently colimit cell growth, (2) acetate only affects cell response to light, and (3) cells keep some level of available nitrogen under starvation, a general multi-resource algal growth kinetics model can be constructed as Eqn. 1, where [P] did not appear because it was provided at concentrations much higher than $K_P$ in nitrogen and light colimitation experiments. The fit of Eqn. S2 gave a value of $N_0$ compatible with zero (p-value = 0.28), therefore, $N_0$ = 0μM was used for fitting Eqn. 1.

Some alternative forms of growth models were also looked at including:

$$\mu = \mu_0 + \mu'_{max} \cdot \frac{L}{K_L+L} \cdot \frac{[N]}{K_N+[N]} \quad \text{(S4)}$$

and $\mu = \mu_0 + \mu'_{max} \cdot min\left(\frac{L}{K_L+L}, \frac{[N]}{K_N+[N]}\right).$ (S5)

Here, a distinct $\mu_0$ was included to address baseline growth. Eqn. S4 and S5 were fitted to the data in a similar way to Eqn. 1. Akaike information criterion (AIC) was used to compare the goodness of fit between models with different number of fitting parameters. The calculated AIC for Eqn. S4 and S5 were -293 and -291 respectively, both higher than -436 obtained for Eqn. 1. This showed that the model presented in Eqn. 1 performed better in fitting as compared to other possible forms.

**Estimation of lake light intensities from clarity measurements**

Light intensities in lakes were estimated from clarity measurements by converting clarity to light extinction coefficient and assuming the surface light intensity. According to the reports from the Citizens Statewide Lake Assessment Program, in the 2020 sampling season, Cayuga Lake (South Shelf Site) and Hemlock Lake (Mid Site) had mean clarity of 2.2meters and 4.4meters respectively. Clarity was measured as Secchi disk depth ($Z_{SD}$), which could be converted to the extinction coefficient of solar radiation (k) by: k=1.7/ $Z_{SD}$ (13, 14). Then the light intensity (I) at certain depth (z) in water could be calculated as $I(z) = I_0 \cdot \exp(-k \cdot z)$, where $I_0$ is the irradiance just underneath water surface. So, at the same depth of 6m under the lake surface (assuming surface light intensity to be 1000PAR), due to the clarity difference, the light intensity in Cayuga Lake was calculated as roughly 0.1PAR, while that in Hemlock Lake, 100PAR.

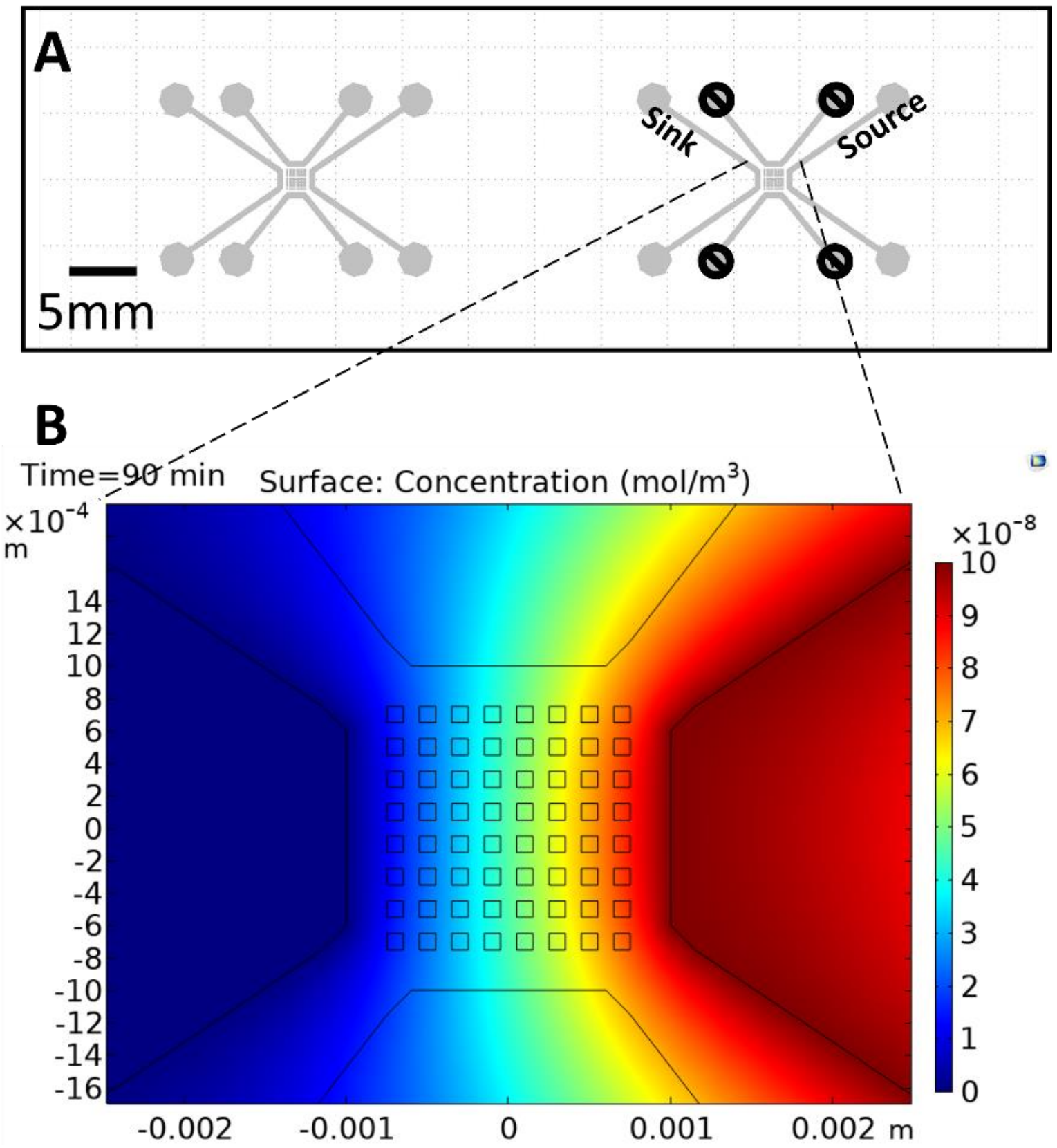

**Fig. S1. Array microhabitat device with chemical gradient generation. (A)** Layout of two devices in parallel on a chip. The device on the right shows the generation of a single chemical gradient by perfusing medium with a target chemical in the source channel and blank medium in the sink channel. The other two side channels are filled will blank medium and plugged. Note that this device has the potential to generate dual chemical gradients, but only one gradient is generated and used in our experiments presented in this manuscript. **(B)** Equilibrium chemical concentration field at time t = 90 min from a COMSOL simulation, where the concentration of the diffusive chemical species was fixed as $1*10^7$ mol/m3 and 0 at the right and left channel respectively, and everywhere else was 0 to start with (at t=0). The diffusion coefficient of the fluorescence dye (fitc) in water was used for direct comparison with experimental results (Fig. 1C).

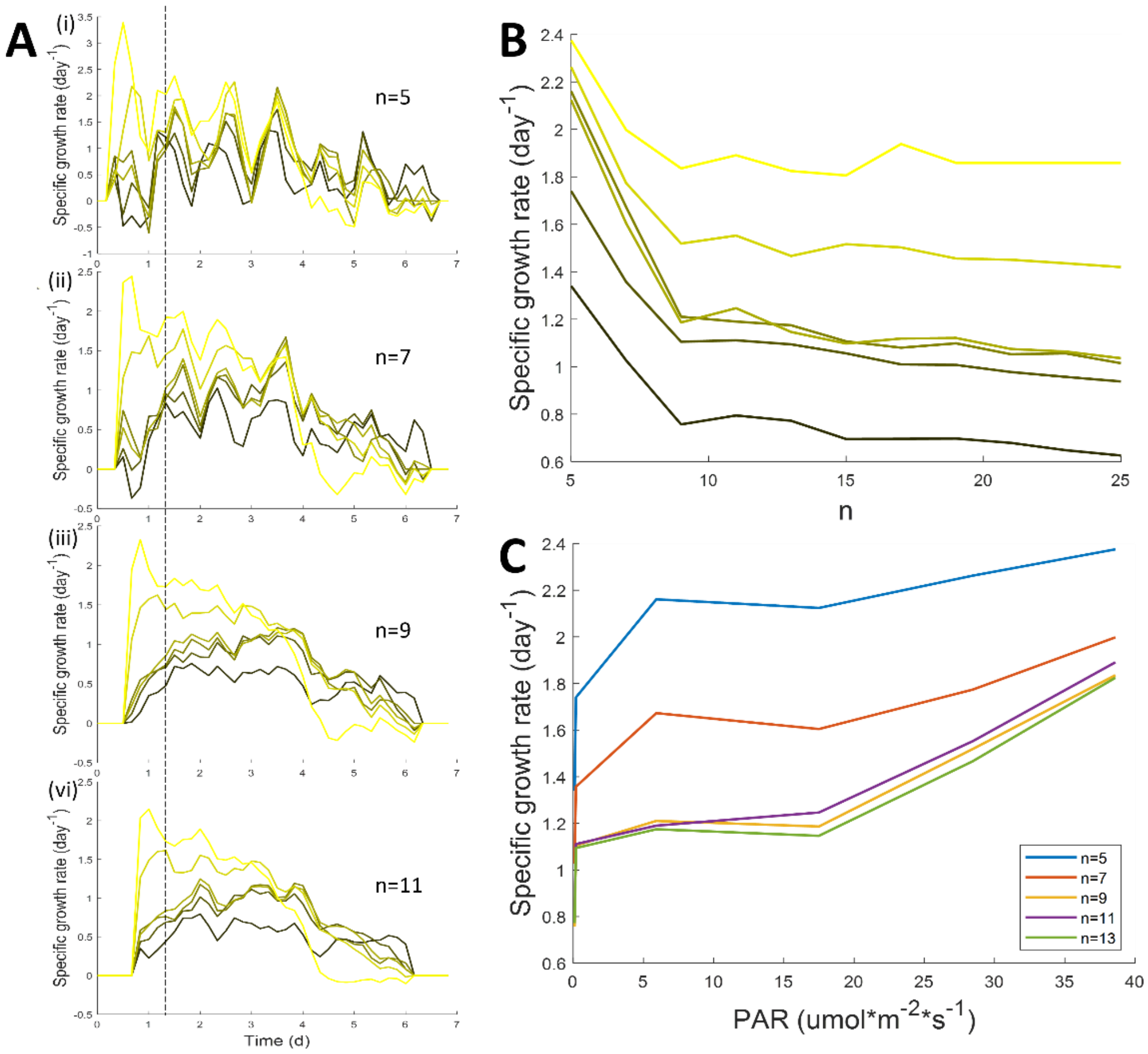

**Fig. S2. Growth rate calculation. (A)** Specific growth rates calculated for growth curves in Fig. 2D, by fitting n consecutive data points around each time point to a straight line and taking the slope. Shown here were n=5, 7, 9, and 11. Colors represent the light intensity conditions (as in Fig. 2D). The dashed line marked day 1.3, after which maximum growth rates were determined. **(B)** Maximum specific growth rates after day 1.3 versus n values. Colors represented the light intensity conditions (as in Fig. 2D). **(C)** Maximum specific growth rates after day 1.3 versus light intensities. Colors represented different n values. n=9 was chosen for final data analysis shown in Fig. 4B.

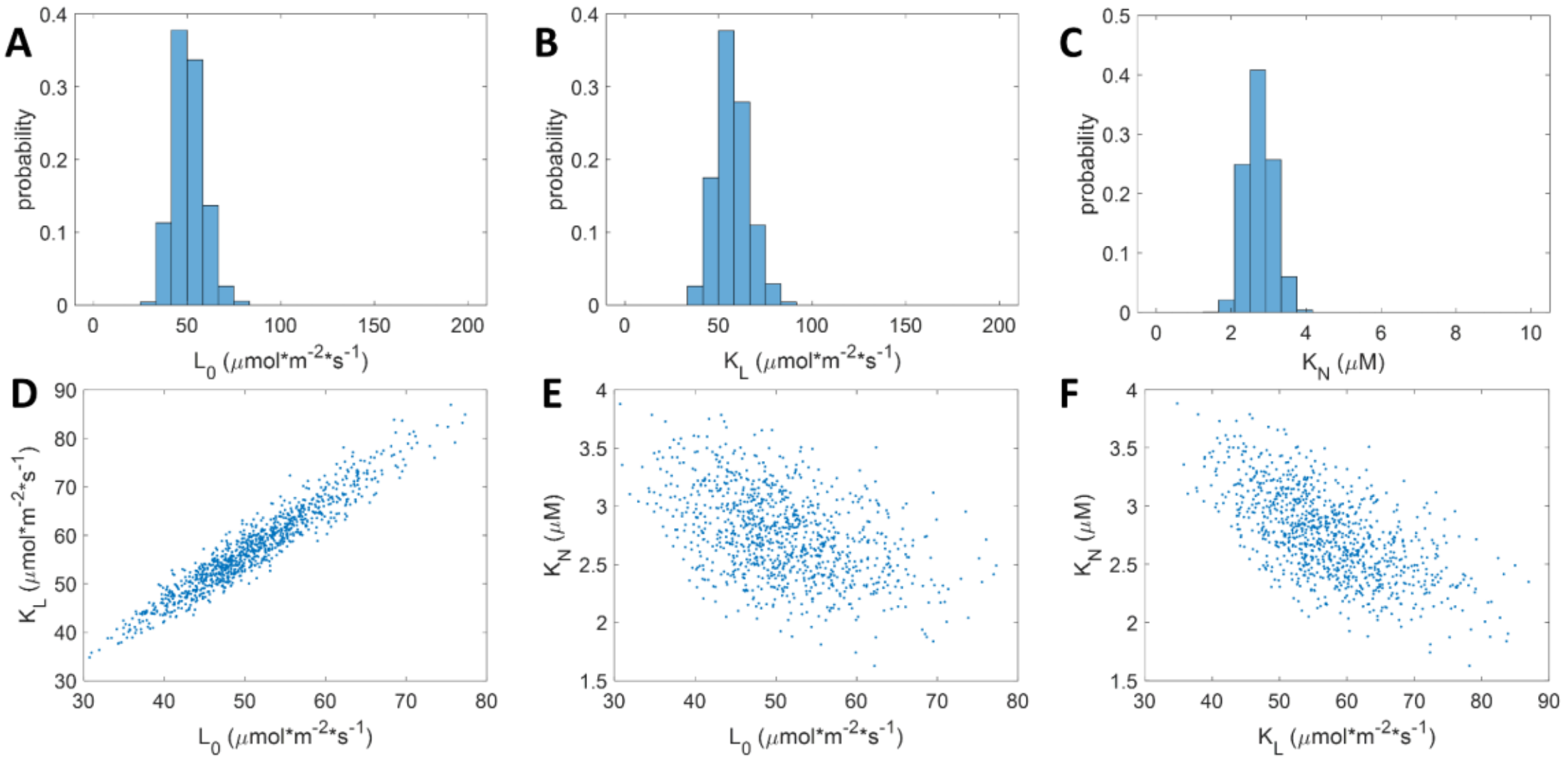

**Fig. S3. Parameters from colimitation model fitting.** Distribution of fitted parameters from the general multiplicative model: **(A)** $L_0$, **(B)** $K_L$, and **(C)** $K_N$. Correlation between parameters: **(D-F)**.

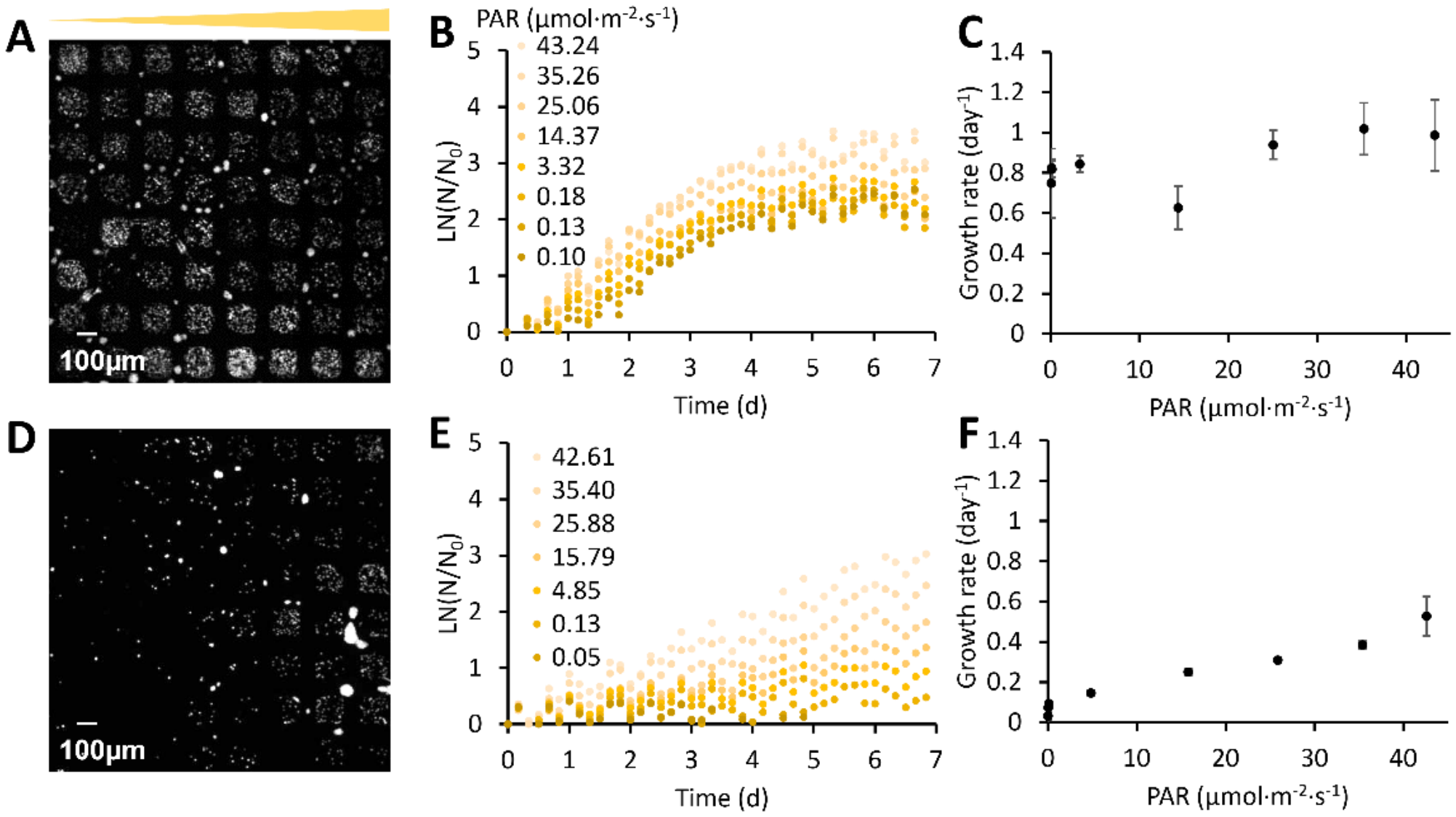

**Fig. S4. Comparing growth of N-starved cells in 10%TAP-N (A-C) and TM-N (D-F). (A)** Fluorescence image of algal cells growing in 10% TAP-N under a light gradient on day 7. Contrast was adjusted for illustration purpose. **(B)** Growth curves of algal cells under different light conditions in the columns in (A). **(C)** Growth rate in 10%TAP-N versus light intensity. **(D)** Fluorescence image of algal cells growing in TM-N under a light gradient on day 7. Contrast was adjusted for illustration purpose. **(E)** Growth curves of algal cells under different light conditions in the columns in (A). **(F)** Growth rate in TM-N versus light intensity. Error bars in (C) and (F) represent standard errors of measurements taken on replicate microhabitats under each light intensity condition. 10%TAP-N and TM-N refer to 10%TAP and TM media with no NH4Cl stock solution added (Materials and Methods), and the nitrogen concentration was 5.3µM in both 10%TAP-N and TM-N.

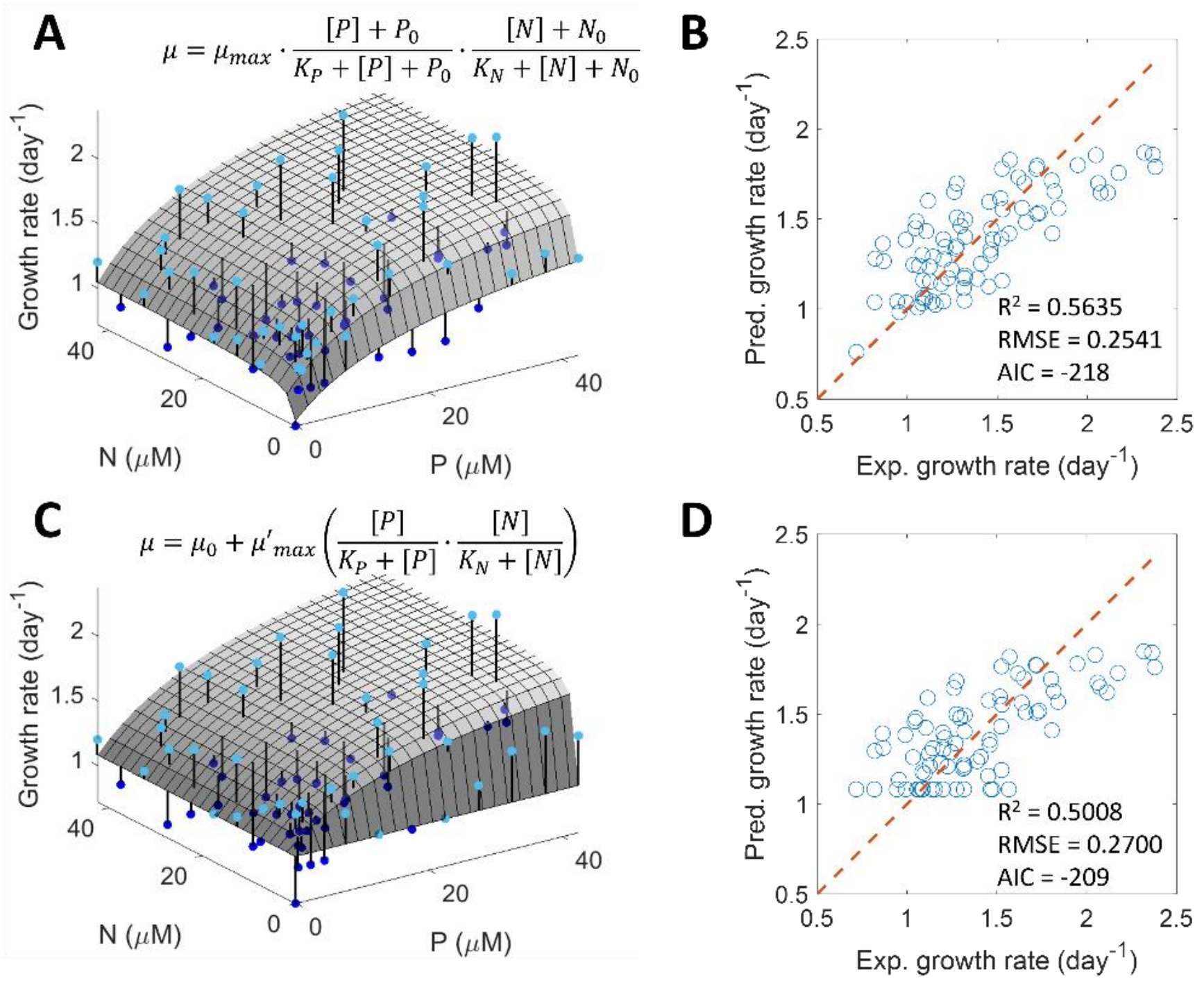

**Fig. S5. Model fitting of phosphorus and nitrogen colimited algal growth. (A,C)** The fitted model (surface) and experimental data (dots) from the average of two replicates, resulted from fitting a model considering nutrient storage (equation S1) (A) and one with only baseline growth (C). **(B,D)** The predicted growth rates versus experimental growth rates showing the goodness of fit for results in (A) and (C) respectively.

**Movie S1 (separate file).** ***C. reinhardtii*** **cells growth in the array microhabitat for 7 days (0-164 hours) under dual light and nitrogen gradients.** Time laps images were taken using fluorescence imaging at 4-hour intervals. Contrast was adjusted for illustration purposes: the intensity (0-255) corresponds to (0-20000) of the grayscale of original images.

## SI References

1. R. Puzanskiy, D. Romanyuk, A. Shavarda, V. Yemelyanov, M. Shishova, A possible molecular mechanism of Chlamydomonas reinhardtii adaptation to different trophic conditions. *Protistology* **15**, 170-189 (2021).
2. Y. Su, Revisiting carbon, nitrogen, and phosphorus metabolisms in microalgae for wastewater treatment. *Sci Total Environ* **762**, 144590 (2021).
3. X. Johnson, J. Alric, Central Carbon Metabolism and Electron Transport in Chlamydomonas reinhardtii: Metabolic Constraints for Carbon Partitioning between Oil and Starch. *Eukaryotic Cell* **12**, 776-793 (2013).
4. S. Chapman, C. Paget, G. Johnson, J.-M. Schwartz, Flux balance analysis reveals acetate metabolism modulates cyclic electron flow and alternative glycolytic pathways in Chlamydomonas reinhardtii. *Frontiers in Plant Science* **6**, 474 (2015).
5. H. C. Huppe, D. H. Turpin, Integration of Carbon and Nitrogen Metabolism in Plant and Algal Cells. *Annual Review of Plant Physiology and Plant Molecular Biology* **45**, 577-607 (1994).
6. X. W. Zhang, Y. M. Zhang, F. Chen, Kinetic models for phycocyanin production by high cell density mixotrophic culture of the microalga Spirulina platensis. *Journal of Industrial Microbiology and Biotechnology* **21**, 283-288 (1998).
7. X. W. Zhang, X. D. Gong, F. Chen, Kinetic models for astaxanthin production by high cell density mixotrophic culture of the microalga Haematococcus pluvialis. *Journal of Industrial Microbiology and Biotechnology* **23**, 691-696 (1999).
8. M. Bekirogullari, I. S. Fragkopoulos, J. K. Pittman, C. Theodoropoulos, Production of lipid-based fuels and chemicals from microalgae: An integrated experimental and model-based optimization study. *Algal Research* **23**, 78-87 (2017).
9. S. J. Yoo, J. H. Kim, J. M. Lee, Dynamic modelling of mixotrophic microalgal photobioreactor systems with time-varying yield coefficient for the lipid consumption. *Bioresource Technol* **162**, 228-235 (2014).
10. V. O. Adesanya, M. P. Davey, S. A. Scott, A. G. Smith, Kinetic modelling of growth and storage molecule production in microalgae under mixotrophic and autotrophic conditions. *Bioresource Technol* **157**, 293-304 (2014).
11. G. M. Figueroa-Torres, J. K. Pittman, C. Theodoropoulos, Kinetic modelling of starch and lipid formation during mixotrophic, nutrient-limited microalgal growth. *Bioresource Technol* **241**, 868-878 (2017).
12. F. Liu, M. Yazdani, B. A. Ahner, M. Wu, An array microhabitat device with dual gradients revealed synergistic roles of nitrogen and phosphorous in the growth of microalgae. *Lab Chip* **20**, 798-805 (2020).

13. H. H. Poole, W. R. G. Atkins, Photo-electric Measurements of Submarine Illumination throughout the Year. *Journal of the Marine Biological Association of the United Kingdom* **16**, 297-324 (1929).

14. B. I. Sherwood, R. G. Gilbert, On the Universality of the Poole and Atkins Secchi Disk-Light Extinction Equation. *Journal of Applied Ecology* **11**, 399-401 (1974).